\documentclass[12pt]{article}
\pdfoutput=1
\usepackage[pdftitle={The Scientific Method},
pdfauthor={B.K. Jennings}, pdfsubject={The Scientific Method},
pdfkeywords={science, philosophy, religion, instrumentalism,
falsification, paradigm}]{hyperref}
\usepackage[superscript]{cite}
\makeatletter
\renewcommand\@biblabel[1]{#1.)}
\makeatother

\begin{document}

\title{The Scientific Method}

\author{B.K.~Jennings\footnote{e-mail: jennings@triumf.ca}\\TRIUMF,
4004 Wesbrook Mall, Vancouver, BC, V6T 2A3}

\maketitle

\begin{abstract}
The nature of the scientific method is controversial with claims that
a single scientific method does not even exist. However the scientific
method does exist. It is the building of logical and self consistent
models to describe nature. The models are constrained by past
observations and judged by their ability to correctly predict new
observations and interesting phenomena. Observations do not prove
models correct or falsify them but rather provide a means to rank
models: models with more ability to predict observations are ranked
higher. The observations must be carefully done and reproducible to
minimize errors. They exist independent of the models but acquire
their meaning from their context within a model. Model assumptions
that do not lead to testable predictions are rejected as unnecessary.
Both observations and models should be peer reviewed for error
control. Consistency with observation and reason places constraints on
all claims to knowledge including religious.
\end{abstract}

\section{Introduction}

The root problem in science is how do we obtain reliable information
on the nature and properties of the real world, whatever the real
world may be. This is essentially the question of epistemology: what
is knowledge and how is it obtained. On the one hand there is the real
world, presumably distinct from the mind. On the other hand, there is
the mind which wants information on the nature and properties of the
real world. The question is how do we connect the two. There is a 
limited number of tools at our disposal to address this issue, only
four: observations, pure thought, innate knowledge, and divine
revelation.  The use of observations to learn about the real world,
done correctly, leads to science. Done incorrectly it leads to
superstition, pseudoscience or abnormal science. The bulk of this
paper is addressed to the question of the use of observation to obtain
knowledge about the nature of the real world. It builds on the
examples and arguments given in ref.~\citen{Jennings} and can be
considered a more formal companion to that work.

\section{Basic Epistemology}

Following Descartes, we have the idea that we can be certain of very
little and start with the position of extreme skepticism. Perhaps much
of what we regard as reality is an illusion: a dream that does not
correspond to reality or the work of a demon deceiving us. As the next
step we rely on Descartes' famous statement\cite{Descartes}: I think
therefore I am. Hence we have two facts. To this we can add a few
others like: I observe. The theorems of pure mathematics
probably\footnote{There are claims that logic is itself empirically
determined and hence uncertain.} also fall into the category of
certain knowledge. This is not knowledge about the real world but
knowledge within the confines of the logical systems of
mathematics. One piece of certain knowledge is that there is very
little other certain knowledge (see discussion in
Sec.~\ref{method}). Even that there is a real world outside the mind
must be held as uncertain: this manuscript may be a figment of your,
the reader's, imagination. This extreme is referred to as solipsism. In
science what are we to make of claims like Isaac Asimov's\cite{Asimov}
that he was happy to have lived at a time when the basic working of
the universe is known?  In theology, what we to make of claims of
proofs of God's existence or non-existence? What does all this imply
about the nature of knowledge?  Some would claim that to be knowledge
a thought must be justified, be true and be
believed\cite{knowledge}. The problem is that, as just argued, very
little is known absolutely to be true. So if knowledge consists only
of things that are known absolutely to be true very little remains of
knowledge outside pure mathematics.

How do we proceed? We make assumptions and test these assumptions. The
assumptions we call models. Knowledge consists of model building and
testing and is always tentative and as argued below frequently
approximate. As discussed in Sec.~\ref{Other} concepts like dog, cat,
and knowledge can also be considered as contributing to knowledge. But
we must for the most part give up the idea of sure and certain
knowledge. Thus we start model building. A first, frequently made
assumption is that there is a reality outside the mind. A second
possible assumption is on the existence of God. Since all nontrivial
knowledge about the world, as a matter of principle, is uncertain
everyone should, at this level, be an agnostic. This concept of
agnosticism is not very useful so instead we can define theist, deist,
agnostic and atheist by the properties of their preferred models. The
preferred model (sometimes called controlling narrative) of theists
would have God actively involved in human affairs or the world, for
the deist a God not involved in human affairs and so on. We proceed
step by step to make assumptions or models and test them. The
assumptions and models always remain tentative and frequently
approximate. Thus Asimov's knowledge is tentative and approximate but
never-the-less comprehensive and useful. Testing procedures are
described in the following sections.

\section{Pure thought, innate knowledge, and divine revelation}
\label{Other}

Before addressing the use of observation to learn about the real world
we turn to the other three tools. The first of these is pure
thought. As a pure abstraction not related to experience, this would
be mathematics which relies only on logic (arguably logic is not
empirically based). As a means to learn about the real world, pure
thought would be synthetic {\em a priori} knowledge as
proposed\cite{Kant} by Kant (b.~1724, d.~1804). Synthetic {\em a
priori} knowledge is nontrivial knowledge about the world obtained
without recourse to observation.  The idea of synthetic {\em a priori}
knowledge was dealt a serious if not mortal blow when Euclidean
geometry, Kant's archetypal example of synthetic {\em a priori}
knowledge, was shown to be not {\em a priori} true with the discovery
of non-Euclidean geometry and to be only approximately true as a
description of the universe's local geometry with the advent of
general relativity. Innate knowledge would be Plato's ideals or forms
and also perhaps logic. Plato's ideals do not exist as such. However
they address a real concern: How does the continuing concept of
``dog'' arise from all the various and changing examples of ``dog'' we
encounter. In the case of a concept like dog, the human mind appears
very good at creating general concepts from a series of specific
instances. The precise definition of a concept like ``dog'' frequently
has ``fuzzy'' edges since it is not clear where the concept like dog
ends and a concept like ``wolf'', ``jackal'' or ``hyena'' begins.  The
concepts are empirically determined by the human mind's marvelous
pattern recognition ability and in science they are judged by their
usefulness in creating and testing models.

Divine revelation, if it exists, could be the most reliable source of
knowledge since it would give information not limited by the
constraints of normal sensory input. However, divine revelation is
communicated using words and language which brings it into the realm
of observations --- either sight or sound. In addition words derive
there meaning from their context even more than observations, and are
less precise than mathematics. Narratives also derive much of their
meaning from their context. It is sometimes unclear if a given
narrative is meant as history, parable, humor or perhaps all three;
the determination depending on its context which may be lost, rather
than purely internal evidence. Words also change their meaning over
time: ``let'' at the time of King James I was used to mean ``prevent''
rather than allow. In addition when divine revelation, in the form of
sacred texts for example, are translated from one language to another
subtlety and meaning are lost. Even copying texts can introduce errors
and ``helpful'' scribes have been known to ``improve''
manuscripts. Moreover many, including deists and atheists (in the
strong sense of the word), would claim that divine revelation does not
even exist.  Even if it does exist, there are many conflicting claims
as to what is the correct divine revelation. Some of the latter
uncertainty is related to the interpretation of words and narratives
as illustrated by the disagreements among the various Christian
denominations. The overriding problem with divine revelation as a
source of knowledge is this: Which, if any, of the many conflicting
purported divine revelations are valid and how do you tell?

Observation and reason provide objective criteria to choose between
the conflicting claims of divine revelations. The argument\cite{Locke}
is essentially that of John Locke(b.~1632, d.~1704): ``Revelation can
not be admitted against the clear evidence of reason.'' and ``If the
boundaries be not set between faith and reason, no enthusiasm or
extravagency in religion can be contradicted.''  Reason should be
supplemented with observation. If a purported divine revelation claims
that when stones are dropped they fall upwards we are safe in
rejecting this as not being a true divine revelation. A similar point
was made by Augustine(b.~354, d.430)\cite{Augustine}:
\begin{quote}
Usually, even a non-Christian knows something about the earth, the
heavens, and the other elements of this world, about the motion and
orbit of the stars and even their size and relative positions, about
the predictable eclipses of the sun and moon, the cycles of the years
and seasons, about the kinds of animals, shrubs, stones, and so forth,
and this knowledge he holds to as being certain from reason and
experience. Now, it is a disgraceful and dangerous thing for an
infidel to hear a Christian, presumably giving the meaning of Holy
Scripture, talking nonsense on these topics; and we should take all
means to prevent such an embarrassing situation, in which people show
up vast ignorance in a Christian and laugh it to scorn. The shame is
not so much that an ignorant individual is derided, but that people
outside the household of the faith think our sacred writers held such
opinions, and, to the great loss of those for whose salvation we toil,
the writers of our Scripture are criticized and rejected as unlearned
men.
\end{quote}
This is consistent with Deuteronomy 18:22\cite{NASB} ``When a prophet
speaks in the name of the Lord, if the thing does not come about or
come true, that is a thing that the Lord has not spoken. The prophet
has spoken it presumptuously: you shall not be afraid of him.'' If
someone makes a claim based on his understanding of divine revelation
that is inconsistent with observation he/she is a false prophet and
the purported divine revelation is invalid. The example of a dropped
stone given above is extreme, but what about the shape of the earth?
Even more pertinent is the nature of planetary motion. Do the planets
circle the earth or the sun? In the early sixteen hundreds, the
Catholic Church and some of its theologians argued strongly that the
claim the earth circled the sun was in direct contradiction with
divine revelation and a threat to Christianity.  Based on observation
and Deuteronomy 18:22 we see that they spoke presumptuously and were
false prophets. Young earth creationists and other people of that ilk
fall into a similar category and speak presumptuously. They are not to
be feared as spokesmen for God. Or in the words of Augustine, what
they doing is ``disgraceful and dangerous''.

Observations can, however, be made consistent with a young age for the
universe through the Omphalos hypothesis\cite{Gosse}, but the cure is
worse than the disease.  The Omphalos hypothesis says that while the
universe is young it has been created with all the appearance of an
old universe. The Omphalos hypothesis, raises significant and
disturbing questions for both science and theology.  Consequently it
was rejected by all sides when first published. In science it is the
question of equivalent models\cite{Jennings} that make the same
predictions for all observations but are internally different. Such
models can only be discriminated between using simplicity or a similar
criteria. Thus the Omphalos hypothesis challenged the idea of the
times that scientific induction could give absolute truth. In
theology, the Omphalos hypothesis raises the question if special
creation can occur without some false history being implied. The
answer\cite{Gosse} from Philip Gosse(b.~1810. d.~1888) is that it
cannot: a creation at a finite point in time, by its very nature,
implies God must implant a false history. This has tended to be
rejected by Christian apologists even more emphatically than an old
universe and with good reason. It portrays God as a deceiver who
planted false histories. Charles Kingsley, author of The Water-Babies
and a friend of Gosse, was asked to review Gosse's book. Refusing, he
wrote to Gosse:
\begin{quote}
    Shall I tell you the truth? It is best. Your book is the first
    that ever made me doubt [the doctrine of absolute creation], and I
    fear it will make hundreds do so. Your book tends to prove this -
    that if we accept the fact of absolute creation, God becomes
    God-the-Sometime-Deceiver. I do not mean merely in the case of
    fossils which pretend to be the bones of dead animals; but in
    \ldots your newly created Adam's navel, you make God tell a lie. It
    is not my reason, but my conscience which revolts here \ldots I
    cannot \ldots believe that God has written on the rocks one enormous
    and superfluous lie for all mankind.  (reproduced from
    Hardin\cite{Hardin} and the Wikipedia\cite{WO}).
\end{quote}
If God made the rocks lie can his words be trusted? Presumably
not\footnote{Even without the Omphalos hypotheses a literal reading of
Genesis challenges God's veracity. Contrary to God's statement in
Genesis 2:17, Adam did not die the day he ate from the tree of the
knowledge of good and evil but rather, like the much maligned snake
correctly stated, he acquired knowledge of good and evil.} and thus in
the end the Omphalos hypothesis would destroy what it sets out to
preserve --- the reliability of the literal interpretation of
Genesis. At one fell swoop it unwittingly questioned and seriously
challenged absolute knowledge in both science and Christian
fundamentalism.

\section{The Scientific Method}
\label{method}

Science is the construction of parsimious, internally consistent
models that can reliably predict future observations. With this
definition the rest of the scientific method follows. It also
sidesteps questions about what the real world ultimately is and even
the nature of truth itself. This definition is not the usual
definition of science. There is nothing here about explaining or
understanding phenomena, nothing about naturalism either
methodological or otherwise nor anything about scientific induction or
falsification. The alternate approaches to science are either
approximations to this approach, partial descriptions that emphasis
one aspect of science or in a few cases simply wrong. The relationship
between other approaches to science and the one given here is the
topic of Sec.~\ref{other}

Observations are any sensory input. At the simplest level they are
direct sensory input --- sight, sound, taste, smell and touch. At the
next level we use instruments to augment our senses, for example
Galileo's telescope. Observations also include controlled experiments
where the experimental arrangements are actively manipulated in order
to isolate different effects and test specific aspects of the
models. The experiments are designed to maximize the information that
can be obtained while eliminating uninteresting and spurious
effects. An extreme example is the ATLAS detector at the Large Hadron
Collider (CERN, Geneva, Switzerland) which is about the size of a five
story building and involves approximately 1500 physicists from 36
countries. As the instrument gets larger and more complex the
observations become more model dependent. A model of the apparatus is
needed or even models of parts of the apparatus. However, even when
the apparatus is as simple as Galileo's telescope, a model is needed
to understand its behavior. One of the attacks against Galileo was
that he was seeing an artifact of the telescope. A not unreasonable
attack since telescopes can ideed add atrifacts, for example,
diffraction rings.  Thus we have the idea that observations are not
freestanding but take on their meaning within the context of a model
or models.

The term ``model'' is carefully chosen. It implies the approximate and
tentative nature of knowledge. It also avoids the ambiguity of the
word ``theory'' which means something different to many people than it
does in science. A model, in the present usage, is anything
constructed in the mind that is used to describe and predict
observations. Knowledge consists two things. The first is the
generation of useful concepts such as dog, cow, or knowledge while
the second is model building and testing. Model building is a creative
activity with relatively few useful guide lines while model testing is
based on a model's efficacy in obtaining a given end and is usually
more algorithmic. In science, models are tested or judged by their
ability to correctly predict observations. In religion, models are
judged by their ability to get the model's adherent to heaven,
enlightenment or nirvana. However in practice this test is usually
replaced by a test based on consistency with a given sacred text or a
given master's teaching. In mathematics, models\footnote{In
mathematics a model would be the set of postulates and theorems that
make up the logical system under study.} are judged purely by their
logical consistency. In capitalism, models are judged by their ability
to generate wealth for the person using the model. In politics it is
the model's ability to get people elected and in auto mechanics it is
the model's ability to aid in car repair. The ability to make correct
predictions is frequently important in judging models in religion (for
example the quote from Deuteronomy given above), capitalism and other
areas.

The idea that we do not have absolute knowledge goes back at least to
Socrates (b.~470?,d.~399 BCE) with the idea of ironic modesty (from
Plato's {\em Apologhma}). Ironic modesty is the claim, attributed to
Socrates, that he was wiser than his contemporaries because he alone
realized how very little he knew. There were also Greek schools of
skeptics (Academic and Pyrrhonian for example) that claimed true
knowledge, especially empirical knowledge, was impossible. Ren\'e
Descartes (b.~1596, d.~1650) in the beginning of his ``Meditations''
makes similar skeptical arguments (before making unwarranted
assumptions to get around the problem). David Hume (b.~1711, d.~ 1776)
pointed out\cite{Hume} that one could never deduce a generalization
valid for all instances from a finite series of observations. Thus he
showed scientific induction is invalid. This was largely ignored by
the scientific community for 200 years. The solution to all these
problems follows from the idea\cite{Popper} of Karl Popper(b.~1902,
d.~1994) that all scientific models (or theories) are tentative. This
should be generalized to the idea that most knowledge is tentative and
approximation. The only exceptions are simple statements like: I am, I
think, I observe and perhaps a few others.

As Hume showed, models can not be proven correct. Despite Popper's
claims\cite{Popper}, in most cases models cannot be falsified
either\cite{Jennings}.  Thus observations neither prove models correct
nor disprove them. What observations do is provide the information
necessary for ranking models.  Old models are displaced by newer
models that are ranked higher based on their ability to correctly
predict observations.  Paraphrasing the United States' gun lobby:
observations do not kill models, models do. This ranking of models is
the first part of the scientific method. 

Comparing models is very much a part of what Popper actually
proposed. For example consider the quote\cite{Popper2} from him:
\begin{quote}
Now there can be little doubt that all these essentialist views stand
in the strongest possible contrast to the methods of modern
science. (I have the empirical sciences in mind, not perhaps pure
mathematics.) First, although in science we do our best to find the
truth, we are conscious of the fact that we can never be sure whether
we have got it. We have learnt in the past, from many disappointments,
that we must not expect finality. And we have learnt not to be
disappointed any longer if our scientific theories are overthrown; for
we can, in most cases, determine with great confidence which of any
two theories is the better one. We can therefore know that we are
making progress; and it is this knowledge that to most of us atones
for the loss of the illusion of finality and certainty. In other
words, we know that our scientific theories must always remain
hypotheses, but that, in many important cases, we can find out whether
or not a new hypothesis is superior to an old one. For if they are
different, then they will lead to different predictions, which can
often be tested experimentally; and on the basis of such a crucial
experiment, we can sometimes find out that the new theory leads to
satisfactory results where the old one breaks down. Thus we can say
that in our search for truth, we have replaced scientific certainty by
scientific progress. And this view of scientific method is
corroborated by the development of science. For science does not
develop by a gradual encyclopaedic accumulation of essential
information, as Aristotle thought) but by a much more revolutionary
method; it progresses by bold ideas, by the advancement of new and
very strange theories (such as the theory that the earth is not flat,
or that 'metrical space' is not flat), and by the overthrow of the old
ones.
\end{quote}

Ranking implies that there must be at least two models to rank. This
is not a problem since there is always the default or null model that
says all outcomes for a given observation are equally possible. Thus
any model that makes some correct predictions but no false ones would
rate higher than the null model. On the other hand a model that makes
only false predictions would rate lower than the null model. We can
improve the null model slightly by having it say that all past
observations are the way they are because that is the way they are and
that any outcome for a future observation is possible. This revised
null model is scientifically equivalent\footnote{see
ref.~\citen{Jennings} for the definition of scientifically
equivalent.} to the ``Because God did it that way'' model.

Thinking in terms of models makes some concepts much clearer. For
example, it is meaningless to ask if a model airplane is a fact,
rather one should ask how accurate the model is. It cannot be one
hundred per cent accurate since then it would be replica not a
model. Even replicas are rarely 100\% accurate.  A similar situation
holds with most models including scientific ones. Asking if Newtonian
mechanics, general relativity or evolution are facts is meaningless,
an error of categories. One should ask how accurately they describe
and predict observations. Rather than fact or not fact, scientific
models are judged on their ability to mirror reality; a mirroring that
can not be exact since models are constructions of the mind while
reality presumably exists outside the mind. This is very much in line
with the skeptical denial of the possibility of absolute
knowledge. However, while models cannot be considered as absolutely
true they can be judged and ranked. Thus we have: Aristotelian
dynamics, Newtonian mechanics, special relativity and general
relativity in order of increasing accuracy of the predictions for the
motion of macroscopic objects. In an absolute sense all these models
are wrong since they do not correctly predict observations on
microscopic systems. In spite of not being absolutely true, all these
models make useful predictions for some range of observations. They
are probably approximations to some unknown model of everything --- at
least that is the hope. Similarly we can rank animals reproducing
after their kind and evolution based on their ability to make
predictions of future observations. Again both these models are useful
although evolution ranks higher based on its predictive ability.  The
ability to rank models is the point missed by Socrates, the skeptics,
Descartes and Hume. Even with naive falsification, there is only a two
level categorization, falsified or not falsified, rather than a
ranking.

The current approach to epistemology, with its emphasis on the
tentative and approximate nature of knowledge, is anti-dogmatic: the
absolute truth of fundamentalists (either scientific or religious) is
a mirage. Even the present description of the scientific method and
the nature of knowledge should be considered tentative and
approximate. Note this circumvents one of the problems with extreme
skepticism. If all knowledge is invalid then even the claim that ``all
knowledge is invalid'' is invalid. Claiming all knowledge is tentative
and approximate does not lead to a similar contradiction.

What properties must a model have in order to be scientific? First we
want to be able to make unambiguous predictions. This implies that the
models should make unique predictions for some set of observations. To
this end we want the models to be internally consistent and logical so
that they do not predict both this and that. At the very least you
need a well-define set of rules that unambiguously lead to testable
predictions.

Parsimony or simplicity is also an important, indeed crucial, property
of models. For a given model it is always possible to make a more
complicated model which makes the same predictions. Comparison with
observation can never discriminate between these models.  This is the
problem the Omphalos hypothesis highlighted. Instead of having the
universe created 6000 years ago we could have it created last Tuesday
or five minutes ago. One can add such assumptions indefinitely: a
universe created last Tuesday, invisible unicorns, flying spaghetti
monsters, etc. However since models are judged by their ability to
make predictions, this gains us nothing. Models should be stripped
down to the minimum assumptions consistent with maintaining their
ability to make predictions. Or rephrasing: The models are simplified
by removing untestable assumptions. If nothing else this makes the
models easier to use. Thus models should be simple, logical and
self-consistent. It is the requirement of parsimony that eliminates
the Omphalos hypothesis and similar untestable hypotheses from
consideration as science.
 
In order to predict future observations it is important to describe
past observations. This is an {\em a posteriori} statement rather than
an {\em a priori} one. While there are few if any hard and fast rules
on how to construct models there is one very useful guideline: The
more one knows about past observations and the patterns in them the
more likely one will be able to construct models that can predict
future observations. Models that incorporate patterns from past
observations typically have more predictive power.  Thus to some
extent, model formation is pattern recognition. Galileo recognized a
pattern in pendulum's period: the period was independent of the
amplitude of the motion. He then used this to predict the periods for
future observations. In this case, the pattern is fairly obvious and
the procedure looks like scientific induction (see
Sec.~\ref{induction}). Most new models in science arise from a
careful, and frequently long, study of past observations. The
scientist has to decide which patterns in the observations are real
and which of the real patterns are important. However describing past
observations and the patterns in them is not enough, the real test of
any scientific model is its ability to predict future observations.

One of the first things one realizes in working with observations is
that it is easy to make mistakes. While an observation is never
``wrong'' it may be misinterpreted. There are some visual impressions
in my eyes but if it is a dog or a cat is a question of
interpretation. It may even be an hallucination and not a real visual
impression. This is also a question of interpretation. Since error and
misinterpretation are ubiquitous, error control is extremely
important. Error control is the distinguishing characteristic between
science and superstition or pseudo-science. There are three main
aspects of error control: care, reproducibility and peer
review. Double blind trails in medicine are an example of the care
required to obtain reliable results. Scientists are frequently
criticized for the care they take to control errors but this is
necessary to prevent science from being overrun with bogus
results. Reproducibility is a major check on both error and
fraud. Repetition is not doing exactly the same experiment again and
again. Rather, the subsequent experiments should be as different as
possible to eliminate common sources of error.

Reproducibility does not mean the scientific method cannot be used to
study historical events because the event cannot be reproduced. The
model that Napoleon died of arsenic poisoning can be tested by looking
for arsenic in a currently existing sample of his hair. Models based
on stomach cancer would not have high levels of arsenic and thus can
be ranked lower if arsenic is found in high concentrations.  The
models of past events can make predictions for future observations
that can be tested. Other historical events, such as the history of
the earth, can be studied similarly.

Next for error control is peer review. This is a simple concept. The
people who know about a topic are peers of the person who did the
original experiment and they look to see if there are any errors. It
is only the peers who would have the knowledge to spot errors.  If you
want to know if a model about sheep farming is reasonable you ask a
sheep farmer. If the model is about Irish history you ask an historian
specializing in Irish history not a sheep farmer (although one could,
in principle, be both sheep farmer and an Irsih historian). If the
model is about nuclear physics you ask a nuclear physicist. Peer
review is this idea applied in a systematic manner.
 
A model airplane by, its very nature, is never an exact replica of the
original. Thus it is always ``wrong''. However we can compare
different airplane models to see which is the most accurate. We can
also ask which is the most useful for a given museum display. This may
well not be the most accurate. The same considerations apply to
scientific models or indeed any models in the mind. By their very
nature they are approximate but the accuracy of different models can
be compared models and ranked. Similarly the most accurate scientific
model is frequently not the most useful for a given
calculation. Quantum theory is not used to calculate planetary motion.

\section{The Pretenders}
\label{other}

In this section we consider the relationship between other proposed
descriptions of the scientific method.

\paragraph{Appeal to Authority:}

Life is not long enough to do all observations and model calculations
by oneself. Thus, inevitably one has to rely on other people who may
be considered authorities. There is nothing wrong with this. The error
occurs when the authority is assumed infallible or nearly so. The late
medieval church fell into this trap with Aristotle. They took his word
as the gospel truth and were caught by surprise when his models of
natural history were shown to be inferior to those developed by
Galileo and his contemporaries.  The British physicists fell into the
same trap with Newton. Taking Newton as their authority rather than
observations they fell behind their competitors on the continent.  In
science careful, reproducible, peer reviewed observation must take
precedent over any other authority in judging models.

An additional comment on the medieval church and Aristotle is in
order. In the twelve hundreds the works of the classical scholars was
rediscovered, in particular the philosophy of Aristotle. Two of the
early participants in this revival where Roger Bacon (b.~1214 d.~ 1294)
and Thomas Aquinas (b.~1225 d.~1574). Aquinas, a Dominican friar,
accepted Aristotle's conclusions and blended them into Christian
philosophy. He consequently was made a saint by the Catholic church in
1323. His emphasis on Aristotle's conclusion contributed to the
conflict between the church and science at the time of Galileo. Roger
Bacon, a Franciscan friar, in contrast to Aquinas emphasized the
empirical aspect of Aristotle's work. Bacon's procedure\cite{Bacon} of
hypothesize, test against experiment, refine and retest is the
precursor of the scientific method and is very similar to the model
building and testing procedure described in this work. Bacon
essentially discovered the scientific method. However, Bacon's ideas
were largely ignored for several generations. It seems that, then as
now, people prefer the convenient answer from an authority even if it
is wrong to a reliable procedure even if it leads to better
knowledge. Therein lies the danger in the appeal to authority.

\paragraph{Scientific Induction:}
\label{induction}
For much of the history of science, from Galileo to Einstein, the
scientific method was considered to be scientific induction.
Scientists began with observations, cautiously proceeded to a
tentative hypothesis describing the observation, progressed to more
secure but still provisional theories, and only in the end achieved,
after a long process of verification, the security of permanent
laws. In 1904 Newton's laws of motion and gravity would have been
considered the prime example of permanent laws derived by
induction. One year later they were no longer considered exact, at
least not by Einstein. Twenty years later, not only had general
relativity shown Newton's laws of gravity to be approximate but
quantum mechanics had undermined much of the philosophy built up
around classical mechanics.  The clock work universe dissolved in
quantum uncertainty. The models in science are frequently built up by
looking for patterns in past observation and extrapolating these to
future observations. The main error in scientific induction is
assuming the deduced law is permanent infallible knowledge.  The
induced patterns are modified based on the new observations and the
procedure repeated. As the quantum replacement of classical mechanics
illustrated no model can ever be assumed absolutely true just because
it has passed all the tests to the present time. In addition the
procedure for going from observations to a model is not always
straight forward or algorithmic. Rather it is a creative
activity. General relativity was not deduced from experiment in any
simple way. Rather it was sheer genius on Einstein's part.

\paragraph{Paradigm Shifts:}

A paradigm\cite{Kuhn}, the set of interlocking assumptions and
methodologies that define a field of study, is also a model. It is the
main model in a field of study and sometimes referred to as the
controlling narrative.  The overarching model, controlling narrative
or equivalently the paradigm provides the foundation for all work in
the field and a common language for discourse. While observations
exist independent of the paradigm, their interpretation depends on the
paradigm. No natural history nor any set of observation can be
interpreted or even usefully discussed without the framework provided
by the paradigm. Paradigms determine the important questions to be
considered.  They can also act to prevent progress when members of a
community are too committed to their current set of models.

Thomas Kuhn (b.~1922 d.~1996) discusses\cite{Kuhn} two kinds of
science --- normal science and extraordinary science. Normal science
is puzzle solving within the context of a paradigm while extraordinary
science is the overthrowing of the paradigm. The present analysis
gives a different view of the distinction although the distinction
itself remains useful. In extraordinary science the model being
challenged is the main model in the field, the paradigm (or model)
which provides the framework for the field. In normal science it is
the subsidiary models, those models that in principle could be derived
from the main model, that are being tested.  Normal science can often
resemble scientific induction with models apparently deduced from
observations. This is especially true in cases like the color of crows
or the period of a pendulum where a regularity is apparent in the
observations.

Extraordinary science generates paradigm shifts, a change in the world
view in the given field. This change of world view is closely related
to Kuhn's incommensurability\cite{Kuhn} (see also
Feyerabend\cite{Feyerabend}). Proponents of the new model and the old
model use a different language and different concepts. They have
different ideas about what the important questions are. Their whole
framework for understanding observations is different. Hence, it may
be difficult to compare the old and new models in detail.  Despite
these dramatic differences, the old models are frequently good
approximations to the new model for a limited range of
observations. By approximation I mean that the new and old models
give nearly identical predictions for some range of observations.

Paradigm shifts do not occur because the old paradigm is falsified nor
because the new paradigm is proven correct. It is not even because the
practitioners have grown tired of the old paradigm. Rather it is
because the new paradigm or model ranks higher than the old paradigm
based on parsimony and its ability to correctly describe past
observations and predict new ones. In the end this is what science
always reduces to.

\paragraph{Falsification:}

Popper claims\cite{Popper} that models can not be proven correct but
they can be proven incorrect.  The problem here is that no theory or
model exists in isolation but is always supported by subsidiary
theories and models.  Thus any test is not just of one model but of
all the subsidiary ones simultaneously.  This objection to
falsification is known as the Duhem-Quine thesis\cite{Curd}. There is
also the question of the interpretation of the observations and the
possibility of modifications to the model to remove the incorrect
prediction. With enough imagination any negative result can be
explained away. The white crow is really a black crow covered with
snow or the sun is reflecting (specular reflection) off the black crow
in such a way it looks white. However, modifying models to circumvent
falsification usually reduces their predictive power. In the case of
creationist natural history and the Omphalos hypothesis it prevents
any meaningful prediction.  While the Omphalos hypothesis can
``explain'' any observation, it is consistent with any possible
observation and has no predictive power.  Contrary to the impression
given by some histories of science, the 1887 Michelson-Morley
experiment did not immediately falsify the ether model. A 1902 high
school physics text book\cite{Gage} defines physics as: ``Physics is
the science which treats of matter and its motion, and of vibrations
in the ether''. Although they developed the mathematics needed for
special relativity neither Lorentz nor Poincar\'e abandoned the
ether. Creative people came up with explanations such as ether
entrainment and the Lorentz-Fitzgerald contraction to explain the
unexpected results. Both explanations reduced the ether's predictive
power.  Ether entrainment say that the ether is dragged along by the
earth. Thus the measured speed of light depends on how strongly the
ether is entrained and accommodates a wide range of values rather than
the single number of the original ether model.  Lorentz-Fitzgerald
contraction adds an extra assumption that does not lead to other new
predictions. Thus it added an assumption for this one type of
observation and the number of predictions per assumption decreased. In
the end, the ether model was eliminated, not directly by observation,
but by Einstein's special theory of relativity which had fewer
assumptions and more predictive power than the competing model.  Using
observations as the criteria, special relativity ranked higher than
the ether model. In general, models are not disproven by observations
but are replaced by newer models with more predictive power.

In cases like the ether or special creation the loss of predictive
power resembles falsification. In many cases falsification can be
considered a useful approximation to the more general procedure of
model testing through predictive power. To some extent the difference
with Popper is one of emphasis.

\paragraph{Natural Explanations/Methodological Naturalism:}

One view is that science is trying to provide natural explanations of
phenomena. Thus we have to define natural and explanation. One
commonly accepted distinction is between natural and
super-natural. Natural is what science has described and super-natural
is what it has not described. In the present context this usage would
be circular since we are trying to describe the scientific
method. When the term ``natural explanation'' is used in science it
is frequently a code for ``explanations'' of the form ``God did it'' are
rejected. 

The related term naturalism has two distinct meanings. One is that the
super-natural is rejected and not dealt with or claimed to be
nonexistent. The other meaning is that all phenomena are treated
equally whether or not they are natural or super-natural. This second
definition has the great advantage that it does not require the
artificial distinction between natural and super-natural. In the
current context, as in marketing, natural and naturalism have mainly a
rhetorical value.  What we want to exclude from science is not the
supernatural but rather assumptions that have no predictive
power. This is the {\em only} reason to reject the ``Because God did
it that way'' type of model.  However, the ``Because God did it that
way'' explanation rarely has predictive power so in practice the
``super-natural explanations'' do end up being rejected. As Laplace
said\cite{Laplace} in response to Napoleon's question on why his books
on celestial mechanics had no reference to God: ``je n'avais pas
besoin de cette hypoth\'ese-l\`a''. He did not say he rejected it
because science is methodological naturalism. That would have been
poor methodology, then or now. The important point is that the
rejection is done {\em a posteriori} not {\em a priori}. The foes of
science have effectively used the methodological naturalism
description of science to score rhetorical points: Scientists have
closed minds, they reject my hypothesis out-of-hand without giving it
a fair hearing. Thus it is important to reject methodological
naturalism as the definition of the scientific method while realizing,
along with Laplace, that something very similar follows from the model
building and testing procedure. However, if a supernatural model makes
clear, unambiguous and precise predictions for observations, by all
means, science must consider it.

The discussions in this paper should not be taken to imply that
science and religion are intrinsically antagonistic. They are
not. Science is the development and testing of models based on
observation. Religion, in its epistemological content, is the
development and testing of models against divine revelation. There is
no reason {\em a priori} that these two approaches should be
conflict. To the theist or deist science is just the studying of Gods
work as made manifest through observation.  The conflict between
science and religion is purely {\em a posteriori}. Some proposed
divine revelations are inconsistent with observation. As discussed
previously we are justified in rejecting such proposed divine
revelations as being false. Thus science is only in conflict with
false religions, actually only with a subset of false religions ---
those that are inconsistent with observation.

``Explanation'' is an equally rhetorically loaded word that is best
avoided. It carries the idea of understanding. An explanation is
something that gives the hearer the impression that he understands the
phenomena. This is a very subjective criterion. Does quantum mechanics
explain or describe quantum entanglement? Do I need the many-worlds
interpretation to ``really'' understand quantum mechanics? Is
everything else just cookbook physics without real understanding?
Does evolution just describe the origin of the species and do we need
intelligent design to ``really'' explain it? These questions cannot be
uniquely answered. They depend on what a person thinks is needed for a
satisfactory explanation. If we take the role of science to be
providing explanations this lack of uniqueness is a serious problem.
Many worlds and supernatural explanations can only be eliminated by
fiat rather than as part of the more general procedure.  The model
building and testing procedure avoids the problem. Nothing is rejected
by fiat. The scientific method is not making explanations but building
models and testing them. Subjective criteria, like what constitutes an
explanation, are avoided. We construct the model and compare it with
other models using observations and simplicity as the criteria. This
may sound cookbook like --- just follow the instructions, but in
science there are only the two criteria: parsimony and comparison
against observation.

Models constructed using parsimony and comparison against observation
are frequently considered to be explanations, especially by observers
who have grown up with them. Newton's law of motion are generally
accepted as explanations of planetary motion in contrast to the
Ptolemaic and Copernican models which are considered only descriptive.
However, it is more precise to say that Newton introduced a higher
level of abstraction using ideas farther removed from the
observations. It is this higher level of abstraction that allows the
models to be taken as explanations. However the role of science is not
to provide explanations, a subjective concept, but to build models.
Providing explanations is seductive but the ultimate insult to a
scientific model is this: It explains everything, but predicts
nothing!

\paragraph{Anything Goes:} 

In areas of science that are undergoing active development we are
faced with incomplete data, wrong data, approximate models and general
confusion. It is like doing a jig-saw puzzle with some pieces missing,
some pieces from a different puzzle and the picture on the box being
wrong in undisclosed ways. Out of this the scientist tries to find
order by constructing models. In the jig-saw analogy he is trying to
construct the picture that should be on the box. The construction of
the new models requires judgment on which data are correct and most
relevant. It requires selective abandonment of old models. For
example, Galileo had to replace the Aristotelian laws of motion in
order to make the Copernican model work. The acceptance of statistical
mechanics was delayed because the physicists of the time assumed
Newtonian mechanics was correct while in the end quantum mechanics was
required. From the general confusion in a developing area it is not
surprising that Paul Feyerabend(b.~1924, d.~1994)
concluded\cite{Feyerabend} that there was no method in science. Hence
the title of his book: ``Against Method: Outline of an Anarchistic
Theory of Knowledge''.  To this extent he is correct: models are not
derived by any logical procedure. Model creation is indeed chaotic
requiring creativity, good judgment and even good taste as to what is
important and relevant. However, above the chaos sit two judges:
predictive power and simplicity. In the end, order emerges from the
chaos by insisting the models be as simple as possible and that they
make predictions that can be tested against future observations.

\paragraph{The EPR view of Science}

Although Lorentz and Poincar\'e developed the mathematics for special
relativity they were unable to accept that the ether was
unnecessary. This crucial insight was made by Einstein who was thus
credited with the discovery of special relativity. In turn, Einstein
helped lay the foundation for quantum mechanics but was unable to
accept its implications. In 1935, Einstein, Podolsky, and Rosen
published a paper\cite{Einstein} (the EPR paper) on quantum mechanics
stressing its incompleteness when judged using classical
concepts. While usually used to judge interpretations of quantum
mechanics it provides an interesting insight into the concept of the
scientific method of that time. This paper preceded the work of
Kuhn\cite{Kuhn} and Popper\cite{Popper} by more than twenty-five
years. In contrast to these later developments in the understanding of
the scientific method, the EPR paper presents a very nineteenth
century view of science.  For example the second paragraph begins
with:
\begin{quote}In attempting to judge the success of a physical theory,
  we may ask ourselves two questions: (1) ``Is the theory correct?''
  and (2) ``Is the description given by the theory complete''
\end{quote}
Non-relativistic quantum mechanics is certainly not correct in an
absolute sense since it does not include relativity. Even quantum
field theory, which is consistent with special relativity, is not
correct in an absolute sense since it is not consistent with general
relativity. Thus the answer to the first question is ``No''.  This is
true of most physical models and will continue to be true until we
have a theory of everything. Even then we will be unable to
demonstrate that the answer is ``Yes'' because all scientific
knowledge is tentative.  In addition to being tentative knowledge is
also approximate\cite{Asimov,Jennings}. Since science is the art of
the reasonable approximation a better question would be: ``Is the
model approximately correct?''.

The EPR paper defines completeness in terms of reality which in turn
requires reality to be defined. The whole emphasis on reality runs
counter to Kuhn's\cite{Kuhn} idea that observations take their meaning
from models. Reality in classical mechanics and quantum mechanics are
inherently different in line with Kuhn and
Feyerabend's\cite{Feyerabend} contention that different paradigms are
incommensurate. There is not a one-to-one correspondence between the
concepts in different models even when they are given the same
names. We can say quantum mechanics is quantum mechanically complete
but not complete according to the concepts of classical
mechanics. There is no reason it should be since the concepts of
reality and completeness are different (incommensurate) in the two
models. If string theory is even approximately correct, reality is
something peculiar happening in some weird number of dimensions. This
is different from reality in either classical or quantum mechanics.

The questions asked in the EPR paper should be compared with the Niels
Bohr's understanding\cite{quantuma}: ``There is no quantum
world. There is only an abstract physical description. It is wrong to
think that the task of physics is to find out how nature is. Physics
concerns what we can say about nature.'' This is a rejection of the
absolutism implied by the EPR questions. In light of the tentative and
approximate nature of physical models and the incommensurate nature of
the concepts in different models, EPR's two questions should be
replaced by one question: ``How much predictive power does the model
have?'' This would destroy much of the force of the paper as an attack
on quantum mechanics. However quantum entanglement, which was
introduced in the EPR paper, is a very interesting and has stimulated
much productive research. .
  
\paragraph{Science and technology}

There is sometimes confusion between what is science and what is
technology. Science is the development of models with predictive
power. Technology is the use of these models to build devices to serve
some function. The development of technology frequently involves model
building and testing using procedures very similar to those used in
science. Thus while the distinction between science and technology is
in principle clear, in practice it is not with some activities
legitimately considered to be both. For example, while Edison was
primarily a technologist, he produced some very interesting science:
for example the Edison effect relating to the current in light bulb
with an additional electrode. Although Edison saw no use for the
effect it eventually lead to the vacuum tube and the cathode ray tube.

The important part of science for technology is its ability to make
correct predictions. When the on switch is pushed we predict that the
TV will turn on. Much science, both pure and applied has to be
correct\footnote{By correct we mean that the scientific models must
make correct predictions for observations.}, in order for that to
happen. Contrary to some claims, the success of technology in building
devices is a validation of science and the predictive power of its
models. Without this predictive power, which is the essence of
science, technological innovation would slow to a barely perceptible
crawl. In return, technology allows science to address interesting new
questions and answer old perplexing questions.

\section{Conclusion}

The scientific method is the building of logical and self consistent
models to describe nature. The models are constrained by past
observations and judged by their ability to correctly predict new
observations and interesting phenomena.  Observations usually do not
prove or falsify models but rather provide a means to rank models. The
observations exist independent of the models but acquire their meaning
from their context within a model. Observations must be carefully done
and reproducible to minimize errors. Models assumptions that do not
lead to testable predictions are rejected as unnecessary.
 
The alternate understandings of science and epistemology have been
proposed at various times in human history. As argued in the previous
section many of these can be considered as approximations to the
current understanding valid for a limited range of situations, much
like classical mechanics can be considered an approximation to quantum
mechanics.  Among the rejected pretenders are appeal to authority,
scientific induction, falsification, paradigm shifts, natural
explanations, methodological naturalism and anything goes. The present
description of science should also be considered tentative and
approximate.

The scientific method does not lead to sure and certain knowledge but
rather to approximate and tentative, but never-the-less useful,
knowledge. This is the best that can be done: General epistemological
arguments, dating back to the ancient Greeks and amplified at various
times since, eliminate {\em all} claims to nontrivial sure and certain
knowledge. In all areas of knowledge, testing against observations is a
powerful filter especially when coupled with predictive power. This
filter is particularly useful in eliminating superstition,
pseudo-science and bogus claims of divine revelation.

ACKNOWLEDGMENTS: B.~Davids is thanked for carefully reading the
manuscript. The Natural Sciences and Engineering Research Council of
Canada is thanked for financial support. TRIUMF receives federal
funding via a contribution agreement through the National Research
Council of Canada.

\end{document}